%% file: main.tex
\documentclass[journal]{IEEEtran}
\usepackage[T1]{fontenc}
\usepackage[pdftex,bookmarks=true,bookmarksopen,bookmarksnumbered,
                colorlinks,
                linkcolor=blue,
                citecolor=blue]{hyperref}

\usepackage{cleveref}
\usepackage{multirow}
\usepackage{graphicx}
\usepackage{array}

\begin{document}

\title{Development of an Undergraduate Quantum Engineering Degree}
\author{\IEEEauthorblockN{A. S. Dzurak, J. Epps, A. Laucht, R. Malaney, A. Morello, H. I. Nurdin, J. J. Pla, A. Saraiva, and C. H. Yang.}
{\IEEEauthorblockA{School of Electrical Engineering and Telecommunications, \\
University of New South Wales,\\ Sydney, New South Wales 2052, Australia.}}}



\maketitle

\begin{abstract}
Quantum technology is exploding. Computing, communication, and sensing are just a few areas likely to see breakthroughs in the next few years. Worldwide, national governments, industries, and universities are moving to create a new class of workforce - the Quantum Engineers. Demand for such engineers is predicted to be in the tens of thousands within a five-year timescale. However, how best to train this next generation of engineers is far from obvious. Quantum mechanics -- long a pillar of traditional physics undergraduate degrees -- must now be merged with traditional engineering offerings. This paper discusses the history, development, and first year of operation of the world's first undergraduate degree in quantum engineering. The main purpose of the paper is to inform the  wider debate, now being held by many institutions worldwide, on how best to formally educate the  Quantum Engineer.
\end{abstract}


\IEEEpeerreviewmaketitle

\section{Introduction}
\IEEEPARstart{S}{ome}
 10 years ago several University of New South Wales, Sydney\footnote{The University of New South Wales (UNSW) has several campuses, the largest one being based in Sydney (UNSW Sydney), where the new quantum engineering program runs. For simplicity, we refer to this campus in this article simply as UNSW. } staff (Andrew Dzurak, Robert Malaney, and Andrea Morello) met to discuss a brand new concept -- the creation of new university courses specifically designed to encourage commencing engineering students to take a ``quantum leap'' into the educational unknown and become ``Quantum Engineers.'' No template existed to guide such a concept:  barriers loomed and the pitfalls aplenty. The anticipated audience was final-year electrical engineering undergraduates and new master's students in electrical engineering and telecommunications. Such students were highly skilled in the fundamentals of engineering but with a background in quantum mechanics which was, at best, a few weeks of a standard Schr\"odinger equation-based introductory-physics course, long lost to the memories of a fun-filled fresher year. They ``never really understood that quantum stuff'' they would later comment, ``but who cares'' they rationalized given ``for the real engineering I am doing  quantum mechanics is not needed that much.''

 The ``carrot'' to entice these students to take the needed leap was the ``whistle in the wind'' of quantum technology \cite{steffen2011quantum}. This new technology promised the world. Quantum computers that were just ``10 years away'' would lead to breakthroughs across many disciplines from miracle new drugs \cite{lanyon2010quantum}, solutions to critical engineering optimization problems that seemed forever unsolvable \cite{farhi2015quantum}, and the direct simulation of the fundamental physics that govern the universe \cite{boghosian1998simulating}, to name but a few exciting issues. Beyond this, quantum communications was already being touted as a ``here and now'' technology \cite{gisin2007quantum}. Unconditionally secure communications were on the verge of being mainstream, and the security issues that plagued the exponentially increasing world of classical communications would soon be forever exiled to history. The gamble the UNSW staff were about to take was - would the promise of  quantum technology be enough for hesitant students to take the jump into the unknown and put at risk their precious weighted average mark?

Fast forward to today. Quantum computers are here in the sense that so-called quantum supremacy over the best classical computers has been delivered \cite{gibney2019hello}. The giants of the commercial world are in the fray \cite{macquarrie2020emerging}, and the world is now truly abuzz that the still-colossal challenges that face the universal quantum computing platform will be overcome soon \cite{debenedictis2020quantum}. Real quantum computers are now only ``5 years away'' state many informed commentators \cite{waters2021goldman}. Quantum communications have indeed taken off; many commercial offerings are available \cite{pittaluga2021600km}, and the technology has been deployed in space on-board the Chinese satellite Micius \cite{gibney2016chinese}. Based on discrete variable photon technology, this low-earth-orbit satellite produces entangled photon pairs, beaming a photon from each pair down to a separate receiver on earth. Thanks to Micius, quantum entanglement has been distributed over thousands of kilometers \cite{yin2017satellitetoground},  quantum-secured networks of large scale are in place \cite{chen2021integrated}, and the global quantum internet is now firmly in sight \cite{yin2020entanglementbased}. The first incarnation of these new communication platforms will connect classical computers worldwide but secure their communication with Quantum Key Distribution \cite{pirandola2020advances}. The worldwide connection of quantum computers will follow. Beyond this, a ``new kid on the block'' has appeared - quantum sensing. Using the heightened sensitivity offered by quantum states, great promise in advanced clocks \cite{smith2019quantizing}, mineral discovery \cite{lautier-gaud2021operating}, enhanced positioning  \cite{duan2021survey}, super-resolution microscopy \cite{healey2020comparison}, single-photon astronomical observation \cite{khabiboulline2019quantumassisted}, and quantum-enhanced imaging \cite{magana-loaiza2019quantum}, are just but a few of the next-generation quantum sensing applications within grasp.

Understanding these dynamics, national governments all over the world are now investing heavily in developing the new workforce that will underpin the emerging quantum economy \cite{csirogrowing,rasanen2021path,quantumcomputing,friedrichs2021analysis}. Commercial organizations are also moving quickly in the same direction \cite{bova2021commercial,hughes2021assessing}. It is the golden era for newly trained engineers -- expert across both traditional engineering offerings \textit{and} quantum technology.  Within this background, the demand for quantum engineering degrees is now palpable and several authoritative papers on the subject of how universities can best plan for the new undergraduate quantum engineering degree have recently appeared in the literature \cite{aiello2021achieving,raino2021quantum,asfaw2021building,perron2021quantum}.

The course offerings on quantum engineering first introduced at UNSW ten years ago have now been taken by over approximately a thousand students. The program has grown substantially, and last year blossomed into the world’s first undergraduate degree in quantum engineering offered by an Engineering Faculty \cite{bachelor}.  This paper outlines the journey to this new degree, the lessons learned, and the challenges still to be faced. Hopefully, it can inform the wider educational discussion now being undertaken by the worldwide university community.


%
%
%
%

\section{Degree Structure and Course Statistics}
The philosophy of the four-year Bachelor of Quantum Engineering intentionally preserves the core structure and attributes of other Bachelor of Engineering programs at UNSW, as follows. In terms of structure, the first year is firmly grounded in fundamental sciences, particularly mathematics and physics, while also introducing students to the engineering design process and electrical circuits. The second year advances analog and digital circuit theory, signals and systems, and skills in parallel with more advanced mathematics. The Fundamentals of Quantum Engineering course, together with the third year of the program provide the foundational courses for each of the key quantum engineering sub-disciplines (see Section III), and then the final year offers elective courses and honours thesis choices for students to specialise further according to their professional interests and needs. Engineering design courses are a common thread throughout the entire degree program, and nearly every course includes a hands-on laboratory program that is either scaffolded to connect closely with the theory in the course, or adopts a problem-based learning approach intended to maximise self-learning of design and complex problem-solving skills. These courses also address the identified need for graduates to be effective in team projects and in communicating technical concepts to a less technical audience \cite{aiello2021achieving}.

The UNSW Bachelor of Quantum Engineering also inherits many courses from a conventional electrical engineering program. The reasons for this are several. Classical microelectronics already operates at a scale where quantum phenomena need to be understood and harnessed. Besides the emergence of band structure in solids -- a quantum phenomenon that can be in principle forgotten once it is understood how it affects charge transport in semiconductors -- we now deal with such small-scale transistors that quantum tunnelling and single-impurity effects \cite{pierre2010singledonor} can have visible consequences. Another reason is that quantum-limited amplifiers are moving from fundamental research into engineering applications \cite{ranjan2020electron}. On the other hand, classical electronics and control systems are of paramount importance for the development of quantum technologies. For instance, large-scale quantum computers employing quantum error correction \cite{fowler2012surface} need Field-Programmable Gate Arrays (FPGAs) for fast feedback \cite{ofek2016extending}, and enormous classical computational resources to interpret the outcome \cite{devitt2014classical}. Virtually all examples of solid-state quantum computers require advanced microelectronic and microwave engineering for the generation \cite{vandijk2019impact,bardin201929,pauka2021cryogenic} and delivery \cite{dehollain2012nanoscale,bejanin2016threedimensional} of the control signals. From the perspective of the emerging quantum industry, a recent study indicates that among a survey of 21 US quantum companies, 95\% of which employed engineers, 71\% employed electrical engineers, a significantly greater proportion than any other type of engineer \cite{fox2020preparing}. The progress of quantum engineering thus hinges upon the availability of a workforce skilled across the broad spectrum of quantum technology and electrical and computer engineering.

Those courses in the degree program without ``quantum'' in their title are regular electrical engineering courses, with two exceptions to date. The first-year Electrical Circuits course, which covers fundamental principles such as the operation of logic gates, operational amplifiers and resistors, capacitors and inductors, also contains a short introduction to qubits. This allows first-year quantum engineering students to begin making a direct connection with their discipline. The second-year Electromagnetic Engineering course covers phenomena in wave propagation such as discrete modes in waveguides, evanescent waves, and reflections from impedance mismatch, which are completely classical. Yet, their mathematical description and physical intuition applies similarly to the quantization of particle states in a confining potential, or quantum tunneling and reflection by a potential barrier. These aspects are strategically emphasized, so that students recognize their ubiquity once they encounter them again in the Fundamentals of Quantum Engineering course. These courses provide a solid basis for the core quantum engineering-specific courses, which are described in more detail in the following sections.

\begin{table}[h!]
\caption{Structure of Bachelor of Quantum Engineering Program at UNSW}
\label{table:data_info}
\begin{center}
\setlength{\tabcolsep}{0.4em} 
{\renewcommand{\arraystretch}{1.5}
\begin{tabular}{|c|m{0.29\linewidth}m{0.29\linewidth}m{0.29\linewidth}|}
\hline
\multirow{3.75}{*}{\rotatebox[origin=c]{90}{Year 1}} & Mathematics 1A & Mathematics 1B & Mathematics 2A \\
& Electrical Circuits & Physics 1A & Physics 1B \\
& Introduction to \newline Engineering Design & Programming \newline Fundamentals  &  ~ \\
\hline
\multirow{4.5}{*}{\rotatebox[origin=c]{90}{Year 2}} & Circuits \& Signals & Mathematics 2B & \hyperref[subsec:fundamentals]{$^A$}Fundamentals of \newline Quantum Engineering\\
& Digital Circuit \newline Design & Analog Electronics & Digital Signal \newline Processing \\
& Electromagnetic \newline Engineering & Engineering Design & \emph{General Education} \\
\hline
\multirow{3.75}{*}{\rotatebox[origin=c]{90}{Year 3}} & \hyperref[subsec:communications]{$^B$}Quantum \newline Communications & Quantum Physics \newline of Devices \& Solids & Electrical Design \newline Proficiency \\
& Electronics & Control Systems & ~\\
& \emph{General Education} & Electrical \newline Engineering Design &  \emph{Elective L3/L4} \\
\hline
\multirow{4.5}{*}{\rotatebox[origin=c]{90}{Year 4}} & Research Thesis A & Research Thesis B & Research Thesis C\\
& Strategic Leadership and Ethics & ~ & \hyperref[subsec:devices]{$^D$}Quantum Devices \& Computers \\
& \emph{Elective L4} & \emph{Elective L4} & \hyperref[subsec:control]{$^C$}\emph{Quantum Control (Elective L4)} \\
\hline
\end{tabular}
}
\end{center}
\end{table}

At UNSW, apart from the newly developed quantum courses, there has also been  success in encouraging electrical engineering students to enter the field via our ``Taste of Research" scholarship and final-year Honours thesis project.
The Taste of Research scholarship program was introduced two decades ago by the Faculty of Engineering, where second or third-year undergraduate students can experience what research entails compared to ordinary lecture-style courses. Many electrical engineering students now choose research topics in the field of quantum engineering for their Taste of Research experience, believing this area is the new rising star.
The Honours thesis topics have also provided a similar platform on drawing students from electrical engineering to quantum engineering.
Both of the above programs have also contributed to quantum-research outcomes at UNSW, and led several students to continue onto  quantum-based PhD programs.

\section{Core Quantum Courses}

\subsection{Fundamentals of Quantum Engineering}
\label{subsec:fundamentals}
The first quantum-centered course encountered by the students is Fundamentals of Quantum Engineering, offered in second year. This course derives from a former postgraduate course, Quantum Devices, and benefits from many of the lessons that were learned while developing such a course. The key insight was to recognize that, while an in-depth understanding of quantum theory is important to gain proficiency in quantum engineering, the use of engineered quantum devices as ``worked examples'' to illustrate quantum phenomena at work is an extremely powerful tool to clarify and demystify the quantum theory.

The first statement -- that quantum theory is essential to quantum engineering -- is debatable. Many attempts exist to teach how to use quantum systems (particularly quantum computers) without knowledge of quantum theory. In our experience, however, the second statement -- that engineered quantum systems greatly help understanding quantum mechanics at work -- is uncontroversial. Many of our students have been enrolled in double degrees with Engineering and Physics, and had taken traditional quantum mechanics within their physics degree prior to the quantum engineering course. They systematically reported that the way they learned quantum mechanics within quantum engineering greatly deepened their understanding of the discipline.

We have found that many students are genuinely curious about the depth of quantum theory, and like asking (and receiving an answer to) the ``why?'' question. This particular course explicitly encourages such intellectual curiosity by rejecting the often-quoted idea that quantum is weird and counter-intuitive (see discussion in next section how the Quantum Communications course approaches this issue). Intuition is the instinctive organisation of past experiences into expectations for what is likely to happen in similar future circumstances. Since our senses have not been developed to detect quantum phenomena, we do not have a natural intuition for them. A key goal of Fundamentals of Quantum Engineering is to provide the students enough exposure to the workings of simple quantum systems that some intuition can begin to develop.

Our student cohort is very diverse. In the original incarnation as a postgraduate Quantum Devices course, the diversity was even more extreme, because many students had completed their undergraduate degree at different universities, in different countries, speaking different languages. Out of acute necessity, we set out to teach quantum engineering while making close-to-zero assumptions on the background knowledge of the students. This naturally resulted in a strong focus on diversity and inclusion, embedded within the teaching rationale of explaining everything from the very beginning, taking nothing for granted, and ensuring that a shared vocabulary is established at every stage.

The method we adopted to expose the students to quantum phenomena was the development of software that simulates the behaviour of increasingly complicated quantum systems. This goes hand-in-hand with a focus on teaching quantum mechanics in a matrix-based form, which lends itself naturally to coding with software packages like Matlab. In 2021 we switched the coding language to Python, in response to its widespread adoption in quantum technologies. The complexity of the systems that can be simulated this way is remarkable, and qualitatively changes the way in which students approach the subject. Traditional textbook examples are defined by the possibility of solving them analytically with manageable difficulty. Now, using numerical software and matrix algebra, students only need to learn the basic methods of writing down a Hamiltonian matrix and asking the computer to calculate its eigenenergies, or writing a time-evolution operator and programming its application to the quantum state of the system. Increasing the system complexity does not reflect in increased difficulty in writing the code. It does, of course, reflect in the computational cost of matrix multiplication as the size of the system grows. Asking the students to calculate the time evolution of an increasing number of coupled qubits is a simple but effective exercise to make them aware of the exponential complexity of quantum systems! We insist that the students write the code themselves, from scratch, rather than use third-party simulation software. This ensures that there are no grey areas in their understanding of how the quantum theory gets translated into computable instructions. The Matlab/Python coding exercises were consistently described as the most valuable and effective learning activity in the students' feedback.

The course  begins with a brief historical exposition of how quantum mechanics emerged as a \emph{necessity} to explain the empirical observation of the photoelectric effect and matter-wave interference. The Heisenberg uncertainty principle is ``derived'' from the properties of Fourier transforms, once wave-particle duality is established. Therefore, from the very beginning, we strive to introduce quantum phenomena as ``the true behaviour of physical systems,'' rather than coating them in an aura of mystery and weirdness.

Matrix-based quantum mechanics is then introduced, using the spin as the working example. Spin precession and spin resonance allow the illustration of many key aspects of quantum engineering, while illustrating them in a system that students find easy to visualize. Students are immediately forced to internalize the dichotomy between ``real'' space and Hilbert space. Several pen-and-paper problems, as well as Python coding exercises, illustrate how to switch back and forth between the two-dimensional Hilbert space with complex coefficients, and the familiar space in which the spin precesses around a magnetic field in a way that reflects classical intuition. Again, this has the explicit purpose of demystifying the presumed weirdness of quantum mechanics. The genuinely nonclassical aspects, such as the spinor and the geometric phase that arises from it, are introduced later in the course, for rigor and completeness.

Once the matrix formalism is established -- both conceptually, and in its practical execution through Python code -- it becomes easy to move to more complex systems where quantum issues are more striking. Entanglement arises naturally and intuitively from the eigenstates of coupled spins. There is no weirdness or mystery in showing that a singlet state cannot be written as a tensor product. Because entanglement is such a key component of quantum advantage, it is essential that its origin and application be demystified and framed as a resource that can be engineered and exploited.

Having established multi-particle statistics, interactions and entanglement, we proceed with a brief discussion of the hydrogen atom, the covalent chemical bond, and the emergence of band structure in semiconductors. The course is concluded with examples of quantum transport and quantum operations in semiconductor devices, and the use of single-electron transistors and quantum point contacts as sensitive electrometers.

\begin{figure}
	\centering
		\includegraphics[width=\columnwidth]{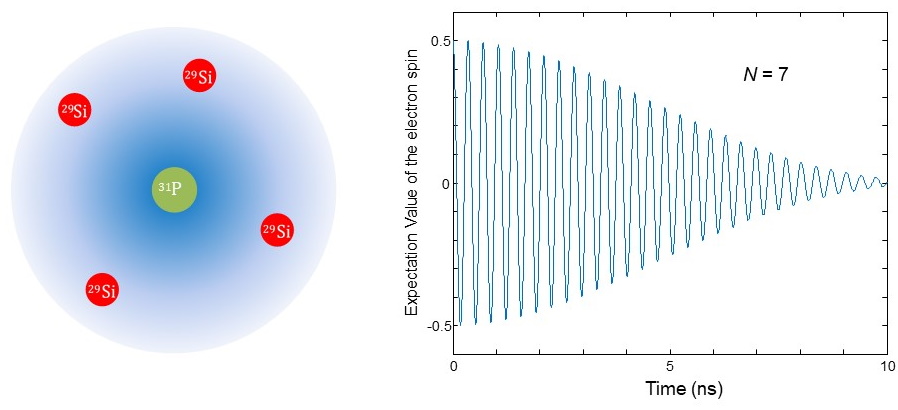}
		\caption{An example assignment for Fundamentals of Quantum Engineering: calculate by ``brute-force'' the time evolution of an electron spin coupled to an increasing number $N$ of nuclear spins. The concrete example shown here is a phosphorus donor coupled to $^{29}$Si nuclei. The Python coding is straightforward to write, involving just the use of tensor products to construct the spin operators in the Hilbert space, and the unitary time-evolution operator to calculate the behavior of the electron. The result illustrates the key phenomenon of decoherence, which arises naturally by randomizing the electron-nuclear coupling strengths and the initial state of the nuclei. It also strikingly illustrates the computational challenge of simulating quantum systems on a classical computer, since the computation becomes very lengthy once $N > 10$. We let the students find this out the hard way, and the lesson learned is hard to forget.}
		\label{assignment}
\end{figure}


\subsection{Quantum Communications}
\label{subsec:communications}
  The Quantum Communications course is offered as a third-year undergraduate course within the new Bachelor of Quantum Engineering program. However, it has been run since  2010 as a master's course, accessible to the traditional undergraduate stream as an elective 4th-year course. Approximately one thousand undergraduate and master’s students have come through the course. Here, we first discuss the development of this course
prior to its inclusion in the new degree.

Perhaps distinct from other emerging quantum technologies, quantum communications exist as a fully deployed technology -- commercial offerings exist \cite{pittaluga2021600km}, real-world case studies are known \cite{bacco2021fullyfledged}, and city-wide deployments have been carried out in several major cities \cite{mehic2020quantum}. This made for a somewhat easier pathway to encourage job-seeking engineering students to enter such an “alien” course.

The cohort of students initially entering the course were predominantly electrical engineering students with a strong background in electromagnetism, signal processing, telecommunications, and classical information theory. It was quickly learned during the first lectures that the best policy moving forward with the course was to assume no quantum mechanic prowess whatsoever within any incoming engineering cohort. We had to start from scratch.

A first challenge  revolved around getting the student cohort to accept the core mystery of quantum mechanics as a fact of nature. Some strategies to achieve this within our program have already been discussed in the previous section  As elucidated by Feynman \cite{feynman1963feynman}, the two-slit interference experiment -- the introduction to many a quantum physics textbook -- goes to the heart of the issue for many of our students.  In the Quantum Communications course, many parallels with classical communications are drawn, and we  attempted to position this mystery to the communication-network context of a single photon encountering a 50-50 beamsplitter that offers the photon two separate paths through the network with equal probability. The notion that a single photon was ``passing'' through the network in multiple paths simultaneously was indeed a hurdle to overcome. Quantum entanglement, and the notion of instantaneous  ``wave collapse'' across dimensions as large as the universe itself, let alone a communication link between Sydney and Melbourne, was another conceptual hurdle. Finally, the notion of measurement changing the very metric being probed seemed to make even less sense to our eager but somewhat bemused cohort.

The resolution to the above issues for the class was the acceptance of the postulates of quantum mechanics, and a frank discussion on the meaning of true -- yet unprovable -- statements. As engineering students, the notion of building a mathematical framework upon solid foundations, known to be true through a century of experimental verification, appeared to hit the comfort zone of the majority of the cohort. Accepting this foundation as the route to mathematical calculations and predictions that would ultimately be tested in real-world deployments was how the course moved forward. This approach, in conjunction with the strategies outlined in the Fundamentals of Quantum Engineering course, seemed to overcome the mystic of quantum mechanics for most of our cohort, allowing them to frame the subject within an engineering context.

Next, we had to decide what mathematical route through the communication journey would be adopted for the course. This would be a two-step process. The first step was to decide whether to adopt the wavefunction framework of Schr\"odinger’s equation and the evolution and collapse of this complex wavefunction to the observable predictions, or to adopt the more abstract formalism offered by time-dependent operators in matrix notation. As the cohort was well versed in matrix operations and eigenvalue analysis, through exposure to such techniques in  other engineering courses, the latter formalism proved more effective. The second step was to decide what photonic technology was to be used as the basis of the course. Would this be the continuous variable approach to quantum communications based on the solid foundation of Maxwell’s theory coupled to a series of Gaussian functions and well-known detection strategies of homodyne and heterodyne detection? At first pass, this appeared to be the obvious strategy, the cohort had a solid foundation in these concepts from their classical communication experience. However, the mathematical complexities of this framework coupled with  quantum mechanics proved to be a bridge too far for many in the class. The more successful route transpired to be the discrete variable single-photon technology and the embodiment of a superposition of polarized states as a pathway to the proverbial qubit -- even though the notion of single-photon communication was alien to the cohort. However, the relative mathematical simplicity of the discrete variable framework allowed for easier inclusion of quantum mechanics, and faster acceptance of the key concepts underpinning the main functional applications of the course.

Based on these decisions, the course moved through its core objectives rather easily. The first of these was a deep understanding of the killer application in quantum communications, Quantum Key Distribution. Roughly half the course was spent on this application. Teaching this involved bringing together not only the conceptual quantum  underpinnings of the application but also its critical dependence on classical communications -- in particular classical error correction as a means of reconciling the secure one-time-pad between transmitter and receiver. This connection back to a subject well known to the cohort appeared particularly motivational. It was a relief, it seemed, to know all that old-fashioned classical stuff will still be needed in the brave new world of quantum communications. Superdense coding, and a full conceptual understanding of entanglement  and its distribution through networks followed thereafter. The third and final component of the course involved an elementary introduction to quantum error correction.

Throughout the course the students were constantly reminded that the same concepts learned here could be immediately applied to quantum computer technology, but that the engineering solutions and applications  being discussed found a ``deployable-now'' home in the communications arena. The knowledge that all they were learning had real-world application today always appeared to be a welcome message to the students, and help defeat any notion that quantum technology was all hype and no action.

\begin{figure}
	\centering
		\includegraphics[width=\columnwidth]{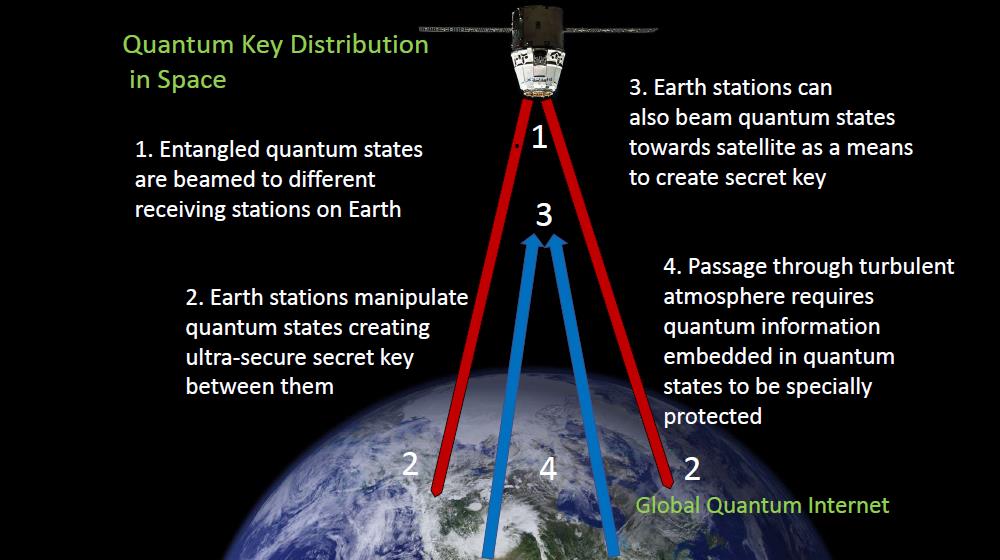}
		\caption{Quantum communications teaching in the degree is based on a thorough understanding of all quantum protocols implemented in space by the Micius collaboration \cite{gibney2016chinese}. That advanced quantum technology has been implemented in the hostile environment of space helped convince the class cohort that quantum engineering is an exciting ``here and now'' field.}
		\label{figQ}
\end{figure}

As stated earlier, approximately one thousand students have traversed the UNSW Quantum Communications course. The feedback from students has been largely positive, many stating that the course was the most interesting one they had encountered in their entire degree. Initially, the numbers were small, some 20 students enrolling in the first year, climbing to a maximum of 120 students a few years ago. The course was given an enormous boost by the launch of the Micius satellite in late 2016 \cite{gibney2016chinese}.  With quantum communications now proven to survive the rigors of space deployment and traversal through the earth's turbulent atmosphere, any notion that quantum communications are not a here-to-stay technology was dispelled. Indeed, since 2016 the course leaned heavily on the implementation of the space-based quantum communications shown in Fig.~\ref{figQ}, pioneered by the Micius collaboration \cite{gibney2016chinese}.
It is safe to say the course gained enormous traction through the exciting results produced by Micius. Coupled with the new space paradigm of emerging low-cost satellite launches, many quantum-enabled satellites are soon to launch \cite{kaltenbaek2021quantum}. This new development in launch availability for space-based quantum devices has  added an exciting new element to the course dynamic.

As stated above, the Quantum Communications course at UNSW is now a core component of the new undergraduate Quantum Engineering degree \cite{bachelor}. A new aspect of the course for this degree is the introduction of a laboratory component. This new laboratory component, (not implemented in the master's degree) is described later
(see Section~\ref{sec:ELEC4605Lab}).

Challenges for the course remain. Providing for seamless co-existence with other courses in the Quantum Engineering degree is a work in progress. Particular issues around avoiding duplication of material yet allowing for each course to be self-contained is a challenge. The trade-off between the consolidation of material already witnessed and avoidance of repetition is a challenge in quantum coursework as it is any other stream. However, given the conceptual leaps involved for the engineering cohort more weight to a revisit of material is currently given. The focus of single-photon technology within the course is also an issue to be resolved.  A true Quantum Engineer emerging from their quantum-focused degree will meet a quantum-technology universe in which discrete variable technology and continuous variable technology are equally prevalent. Additional course offerings that meet this real-world issue will likely be needed as the degree matures.

\subsection{Quantum Control}
\label{subsec:control}
A  new course dedicated to control-theoretic aspects of quantum engineering forms part of the offering within our degree.
A major consideration in the design of this course was the need to minimize  barriers to entry. We assumed the students taking the course would be well-prepared fourth-year   students who have completed a quantum mechanics course and a first electrical engineering control systems course with at least a state-space component.  We also assumed the students would possess standard knowledge of topics in undergraduate level mathematics such as multivariable calculus, linear algebra, complex analysis, and probability and statistics. It is the case that traditional electrical engineering students at UNSW would already be equipped with this knowledge by the time they complete their third year. Given this background, the course has been  designed to be as self-contained as possible. Additional elements of quantum physics (e.g., open quantum systems), probability theory and control theory are weaved into the course to fill remaining knowledge gaps.

Another important feature in the course design is that it approaches quantum control from an engineering viewpoint and builds upon well-established electrical engineering (systems and control) traditions in optimal control \cite{fleming1975deterministic,brogan1991modern}, and stochastic control \cite{fleming1975deterministic,bagchi1994optimal}. The approach adopted promotes a unified view of classical stochastic control and quantum feedback control, with the latter being viewed as a non-commutative generalization of the former to dynamical systems obeying the laws of quantum physics. This approach is in the spirit of the  pioneering work of Belavkin in quantum filtering and feedback control, see, e.g., \cite{belavkin1983theory,belavkin1988nondemolition,bouten2008separation}, which generalized stochastic control theory to the quantum setting.

In terms of coverage, the quantum control course  includes some of the materials proposed in Modules 8, 9 and 10 in \cite{asfaw2021building} that cover topics on Hamiltonians and time evolution, dynamics with time-varying Hamiltonians and open quantum systems, respectively, but presented at a  general level rather than delving into the details of specific physical settings. Beyond this, it   includes optimal open-loop and feedback control of quantum systems. Basic elements of modern probability theory \cite{billingsley1986probability} and classical stochastic control theory \cite{fleming1975deterministic,bagchi1994optimal} are gently introduced as a pre-cursor to the quantum extension. This equips students with a broad awareness of both classical and quantum control theory. Starting with the classical control theory has the benefit of first introducing key concepts in a familiar classical (non-quantum) setting that students would be comfortable with, motivated by the engineering problems that they were developed to solve. The treatment of quantum feedback control  then proceeds along an approach that is standard in the systems and control context and utilizes the close connection and analogies between classical stochastic control and quantum feedback control. This approach is the well-established route of separating the feedback control problem into a stochastic filtering problem and a full-information control problem, known colloquially as the separation principle \cite{fleming1975deterministic,bouten2008separation}. Analogies with the classical setting treated earlier in the course can be drawn upon to treat the quantum setting with minimum additional technicalities. The approach  emphasizes properties of quantum models for quantum feedback control that enable  analogies from the classical theory to be carried over to the quantum setting.

\subsection{Quantum Devices and Computers}
\label{subsec:devices}
By the fourth and final year of study of our program, students have established  foundational knowledge on quantum devices through their courses on Fundamentals of Quantum Engineering and Quantum Physics of Solids and Devices. Having already formed an intuition for quantum mechanics and developed both the analytical and numerical skills to simulate quantum systems, they are ready to explore one of the technologies synonymous with the second quantum revolution, quantum computers. The course Quantum Devices and Computers aims to provide students with a comprehensive understanding of quantum computers from both a software and a hardware perspective, covering the basics of quantum gates and algorithms and examining several platforms for its implementation.

A central theme early in the course is how noise affects real quantum systems, a critical concept for understanding the challenges involved in building a quantum computer and in evaluating candidate physical platforms. We have found the filter function approach to understanding noise and dynamical decoupling \cite{biercuk2011dynamical} as an extremely effective tool for teaching this topic to our students. We describe decoherence as a noise sampling problem in the frequency domain, a concept that sits comfortably with the students who by this point in the program are well-versed in Fourier and Laplace transforms and filter design, knowledge developed in their second-year courses on Circuits and Signals and Digital Signal Processing. Noise spectroscopy and magnetometry are briefly covered to ground the filter theory in practical examples.

The course closely follows the style of Nielsen and Chuang \cite{nielsen2000quantum} to introduce quantum gates and the concept of universal quantum computing. We take a system-agnostic approach in the presentation of this section, opting to investigate physical platforms in the second half of the course. A few key quantum algorithms are explored, including the Quantum Fourier Transform (QFT), the Grover search algorithm, and quantum simulation. These general examples cover a wide class of algorithms and provide enough depth for students to understand how to ``program'' a quantum computer and recognize the power and limitations of the technology. Students are given an assignment which typically involves simulating a quantum algorithm using Matlab or Python. For example, students might be asked to solve the dynamics of a particle in a one-dimensional potential using a quantum circuit – a problem that combines both the QFT and quantum simulation algorithms. A small part of this section is also devoted to compiling and running an algorithm on the IBM Quantum Experience cloud quantum computer \cite{ibm}. This demonstration situates the theory in a real tangible example and shows students the practical limitations of current generation quantum computers.

The quantum harmonic oscillator (QHO) is an important topic that permeates the remainder of the course. In addition to being one of the most fundamental concepts in quantum mechanics, the physics of the QHO can be found in each of the platforms for quantum computation that are covered: semiconductor-based devices, photons/quantum optics, and superconducting quantum electronics. When exploring each physical system, the link to the QHO is made apparent: for semiconductor qubits the confining potential of electrons is parabolic and wavefunctions are given by the Fock-Darwin states, photons are excitations in a mode of an electromagnetic field, and superconducting qubits are electromagnetic modes with an anharmonic energy spectrum. Linking the seemingly very different systems back to the same underlying physics of the QHO establishes a sense of familiarity. The course also demonstrates the components needed in each physical system to implement single-qubit gates and two-qubit gates, which provides students with an understanding of the hardware requirements for building a quantum computer.

Tutorials are run throughout the course to allow the students an opportunity to test their comprehension by solving a set of problems specific to each topic. In addition, a hands-on experimental laboratory (see Section~\ref{sec:ELEC4605Lab}) is operated simultaneously with the lectures and tutorials and links to the sections on noise and quantum optics, with the latter providing an opportunity for the students to demonstrate the general concepts of single qubit gates and entanglement. These two components together with the quantum algorithm programming assignment are critical for developing a deep level of learning and rate highly in student course evaluation surveys.

\subsection{Developing Experimental Laboratories}\label{sec:ELEC4605Lab}
For a comprehensive education, the theoretical skills conveyed in the lectures and tutorials need to be augmented with practical skills, obtained through hands-on exercises and laboratories.

\begin{figure}
	\centering
		\includegraphics[width=\columnwidth]{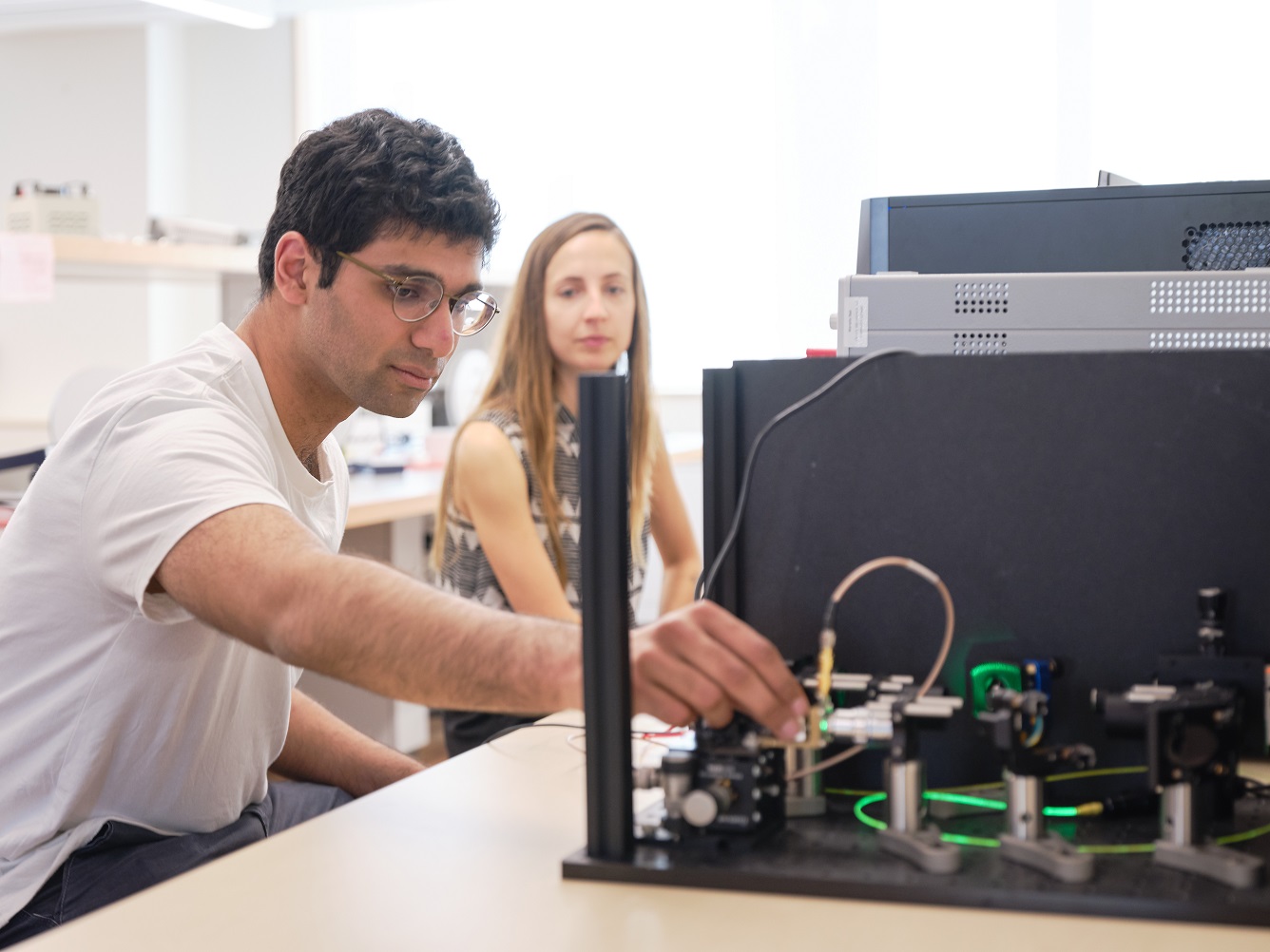}
		\caption{Two engineering students working on the NV center teaching-laboratory experiment. The laboratory components of the  degree are critical in consolidating the quantum mechanical concepts introduced in lectures.}
		\label{fig4605}
\end{figure}

The matrix formalism that is used to introduce quantum mechanics in Fundamentals  of  Quantum  Engineering, Quantum Communications and Quantum Physics of Devices \& Solids lends itself perfectly to performing numerical simulations in software packages like Matlab or Python. A qubit state is simply depicted as a vector, a Hamiltonian as a matrix, and time evolution operators can be obtained by matrix-exponentiation -- all simple linear algebra operations performed by the computer. Going through many examples and requiring students to code up their own solutions in homework tasks and tutorials, the courses provide students with a useful toolbox of codes and techniques that they can later extend and apply to simulate research questions. Examples of publications where the theoretical simulations are based on extended course codes include ~\cite{kalra2014robust,laucht2015electrically,laucht2016breaking,asaad2020coherent}.

For Quantum Devices \& Computers we have implemented and developed hands-on practical teaching laboratories. Half of the teaching laboratories are based on an educational kit sold by the company Qubitekk~\cite{Qubitekk}. Based on a 405~nm laser and a non-linear down-converter, the kit allows students to generate pairs of entangled photons to learn about measurement bases, the photon statistics of non-classical light sources, the indistinguishability of photons, and polarization entangled pairs of photons. In the other half of the laboratories, we convey experimental aspects of qubit operation, by letting the students implement control of an ensemble of NV centers in diamond using magnetic resonance. Here the students learn about Rabi oscillations, two-axis control on the Bloch sphere, coherence times, and dynamical decoupling, having to code and tune up the pulse sequences themselves.

The main technical challenge for the NV center experiment was to take a fragile quantum experiment out of a research laboratory and put it in front of students such that it produces the desired results consistently, in a short time period, and in an ill-controlled environment. While several spin control setups are now commercially available~\cite{spinedu,ciqtek,qunv,spinq}, we designed and built our own setup and made it robust to background light, vibrations, and movement of cables using lock-in amplification techniques~\cite{sewani2020coherent,yang2021observing}.
The main educational challenge for both experiments was to convey the necessary specialist knowledge to the students that is required to run the experiments. While the overarching concepts are covered by the lectures, the intricate details (e.g., quantum properties of the NV centers in diamond) and the measurement techniques are not part of the lectures. To overcome this, we found it helpful to have introductory mini-lectures at the beginning of each teaching laboratory that cover the relevant details. These were very well received by the students and partly moved online to save time.

For the Quantum Communications course, it was decided that although the course was based largely around single-photon technology, the use of weak laser pulses as the core signal carrier for laboratory purposes would be best. The relative ease-of-use, set-up, and lower cost of maintenance of simple lasers relative to single-photon sources was critical in this decision. This laser setup can be used to demonstrate  the important principles of the BB84 Quantum Key Distribution protocol and can be set up using available optics packages \cite{quantumcryptography}.  The demonstration allows for full-encoding and decoding of polarized light, full-blown classical reconciliation and privacy amplification, and the delivery of a final one-time pad. It even allows for interception attacks by a malicious eavesdropper.  Due to the use of a many-photon pulse of laser light, this setup does not allow for an actual secured secret key. However, it was decided that a better understanding of the conceptual underpinnings of the protocol, outweighed the lack of a formally secured key.

\section{Industry Interest}

With commercially valuable applications around the corner, the quantum industry is attracting significant investment. Start-ups are raising millions of dollars in private capital, large corporations are investing in quantum readiness and governments are subsidizing the creation of competitive ecosystems worldwide. Staffing these new companies is a challenge, and competition for talent is fierce.

Part of the difficulty faced by Industry is the lack of a single well-defined degree that immediately guarantees the employer that the candidate has the skills that are needed for the job. For most tasks, currently, it is necessary to hire scientists with either years of proven experience or some research postgraduate degree, such as a PhD.

Entry-level positions are extremely hard to fill. This presents several problems for the competitiveness of quantum companies when compared to other deep-technology industries. The backbone of a quantum company is a well-integrated team, which is easiest to achieve by hiring young talent and offering incentives (such as company shares) and a solid career progression. Hiring more experienced engineers strains the budget of a company in its early stages, in which research and development are intensive and the exact point of effective commercialization might be unclear.

An undergraduate-level quantum engineering program, such as that described here, addresses all these issues. The professional formation is typically achieved in half the time compared to the combination of a Bachelor of Engineering (or other disciplines) and a PhD. Graduates from formal quantum engineering programs will join companies straight out of college, receiving better compensation for their work and creating ties to their company early on. As they progress on the job, their salaries grow commensurately, which makes for a better defined flow of funds for the company and ties the personnel expenses to the company progress -- an attractive feature for investors.

The supply of bachelor-level quantum engineers will also lower the barrier for potential users to hire and develop internal quantum applications programs. The end-user will typically not be interested in developing their own quantum hardware and services, but only in judiciously consuming quantum technologies -- a task easily performed by one or a few quantum engineers.

\section{Global perspective}

Quantum science and technology form a global market. However, different countries  possess unique educational structures and have varying demands regarding a quantum workforce. These differences may affect the strategy for the formation of the quantum engineering workforce.
Thus far, quantum-based companies have successfully built their staff by attracting talent from all around the world. This model is, however, unsustainable moving forward. Incentives for expert workers to move countries is costly and often requires offering long-term contracts, which might not be feasible for pure-play startups.

A quantum-ready workforce is also an important instrument for sustained sovereignty. A locally trained cohort of quantum engineers allows nations to protect their existing technology, and staff their governmental and defense institutions with quantum-literate personnel. Here in Australia, the quantum industry is in its infancy. This is predicted, however, to change significantly in the next years, as predicted by CSIRO in its \emph{Growing Australia's Quantum Technology Industry} report~\cite{csirogrowing}. However, globally, the quantum industry is well beyond its infant stages.

\label{GlobalCurriculum}

An undergraduate degree needs to provide the professional skills that are desirable by industry at an international level. Quantum engineering, being a newly amalgamated discipline, lacks a consensual definition.

An analysis of the quantum industry worldwide may provide some insight into the proficiency required by employers in this field. In a recent survey, 57 companies provided information about the roles they were offering or planning to offer in the near future. At the time of the survey, no undergraduate degree in Quantum Engineering was offered anywhere in the world other than UNSW, and all job titles referred to traditional fields in Engineering and Science, such as Test and Measurement Engineer, Control Systems Engineer and Experimental Physicist. An analysis of the responses on the survey~\cite{hughes2021assessing} may be used to deduce what skills are most sought after by Industry in the hiring of a Quantum Engineer.

Among all the specific skills required for each of the roles, there was a clear correlation between three subsets, which we will call A, B, and C for clarity. The largest subset (A) helps identify a potential curriculum for Quantum Engineering. These skills are Cryogenic Components/Systems; Mechanical Assembly; Microwave Circuit Design; Electronics; Analog Circuit Design; Digital Circuit Design; Circuit or System Testing; Device Fabrication or Process Integration; Device Testing \& Characterization; Control Theory
Device Physics; Noise Measurement and Analysis; Quantum Sensor Physics; Photonics Circuit Design; and
Quantum Photonic/Laser Physics.

Two other skills -- Condensed Matter Physics and
Modeling or Simulation --  were strongly correlated with both subset A and B. Subset B also consisted of Quantum Science (Chemistry, Physics, etc.); Systems Architecture; Error Correction; Theoretical Mathematics or Statistics; Applications Design; Computer Science \& Engineering; AI/ML Algorithm Development; Quantum Algorithm Development; Database Design; Compiler Development; and Software Development. We interpret this latter subset as a Science-oriented potential degree, which might combine skills in Physics, Chemistry, Computer Science and Mathematics.

We also note the existence of a clearly defined subset C, which relates to the quantum industry demand for quantum-literate business experts. While the quantum-specific skills probably are insufficient to warrant the creation of an undergraduate degree associated with this discipline, there is a clear call for  introductory quantum technology education for business professionals.


Efforts to define quantum-specific degrees around the world are reflections of the educational landscape in each country.

The most common offering nowadays is a Quantum Technology degree at the Master's level~\cite{hughes2021assessing}. These degrees typically assume the student has an undergraduate degree in one of the core disciplines surrounding the skills set described in above, such as a Bachelor in Physics or Electrical Engineering.
Short certifications are also a commonly adopted approach, especially in the contexts of quantum programming languages and for quantum finances.




\section{Conclusion}
In this article, we have discussed the development of an undergraduate degree program in quantum engineering - aimed principally at electrical engineering students. This program, which was established over several years, is now running at the University of New South Wales, Sydney, Australia. From it, the world's first batch of undergraduate degree-qualified quantum engineers can be expected in the next few years. We hope the material provided here can inform the current worldwide discussion on the development of undergraduate quantum engineering degrees. In order to  formally train the large number of quantum engineers widely predicted to be high demand in the near future, such degrees will proliferate.




\input{main.bbl}


%



\end{document}